\documentstyle[12pt,a4]{article}
\input epsf
\global\arraycolsep=2pt

\def\spose#1{\hbox to 0pt{#1\hss}}
\def\lsim{\mathrel{\spose{\lower 3pt\hbox{$\mathchar"218$}}
 \raise 2.0pt\hbox{$\mathchar"13C$}}}
\def\gsim{\mathrel{\spose{\lower 3pt\hbox{$\mathchar"218$}}
 \raise 2.0pt\hbox{$\mathchar"13E$}}}

\begin{document}

\begin{titlepage}

\begin{flushright}
CERN-TH/95-255\\
hep-ph/9603414
\end{flushright}

\vspace{0.5cm}

\begin{center}
\Large\bf Improved Bounds for the Slope and Curvature\\
of $\bar B\to D^{(*)}\ell\,\bar\nu$ Form Factors
\end{center}

\vspace{1.0cm}

\begin{center}
Irinel Caprini\\
{\sl Institute of Atomic Physics, Bucharest, POB MG-6, Romania}\\
\vspace{0.5cm}
and\\
\vspace{0.5cm}
Matthias Neubert\\
{\sl Theory Division, CERN, CH-1211 Geneva 23, Switzerland}
\end{center}

\vspace{1.2cm}

\begin{abstract}
We derive a theoretically allowed domain for the slope
$\widehat\varrho^2$ and curvature $\widehat c$ of the physical form
factor appearing in the decay $\bar B\to D^*\ell\,\bar\nu$. Using
heavy-quark symmetry, we relate this function to a particular $\bar B
\to D$ form factor free of ground-state $B_c$ poles below the
threshold for $B D$ production, for which almost model-independent
constraints are derived from QCD using unitarity and analyticity. Our
results are of interest for the extraction of $|V_{cb}|$ from the
recoil spectrum in exclusive semileptonic $B$ decays. Making
conservative estimates of the theoretical uncertainties, we find (up
to $1/m_Q$ corrections) $\widehat\varrho^2<1.11$ and $\widehat
c\simeq 0.66\widehat\varrho^2-0.11$. We also derive the corresponding
bounds for the form factor in the decay $\bar B\to D\,\ell\,\bar\nu$.
\end{abstract}

\vspace{1.0cm}

\centerline{(Submitted to Physics Letters B)}

\vspace{2.0cm}

\noindent
CERN-TH/95-255\\
December 1995

\end{titlepage}

\section{Introduction}

One of the most accurate methods of extracting the parameter
$|V_{cb}|$ of the Cabibbo-Kobaya\-shi-Maskawa matrix is based on the
study of the exclusive semileptonic decay $\bar B\to
D^*\ell\,\bar\nu$. Heavy-quark symmetry (for a review see
Ref.~\cite{review}) can be employed to eliminate, to a large extent,
hadronic uncertainties from the theoretical description of this
process \cite{Volo}--\cite{Vcb}. The analysis consists in measuring
the recoil spectrum, i.e.\ the distribution in the kinematic variable
\begin{equation}\label{wdef}
   w = v_B\cdot v_{D^*} = {E_{D^*}\over m_{D^*}}
   = {m_B^2 + m_{D^*}^2 - q^2\over 2 m_B m_{D^*}} \,,
\end{equation}
which is the product of the four-velocities of the mesons. Here
$E_{D^*}$ denotes the recoil energy of the $D^*$ meson in the parent
rest frame, and $q^2=(p_B-p_{D^*})^2$ is the momentum transfer
squared. The differential decay rate is given by \cite{Vcb}
\begin{eqnarray}\label{BDrate}
   {\mbox{d}\Gamma(\bar B\to D^*\ell\,\bar\nu)\over\mbox{d}w}
   &=& {G_F^2\over 48\pi^3}\,(m_B-m_{D^*})^2\,m_{D^*}^3
    \sqrt{w^2-1}\,(w+1)^2 \nonumber\\
   &&\mbox{}\times \bigg[ 1 + {4w\over w+1}\,
    {m_B^2-2w\,m_B m_{D^*} + m_{D^*}^2\over(m_B - m_{D^*})^2}
    \bigg]\,|V_{cb}|^2\,{\cal F}^2(w) \,,
\end{eqnarray}
where ${\cal F}(w)$ is the (suitably defined) hadronic form factor
for the decay. Apart from symmetry-breaking corrections that can be
calculated using the heavy-quark effective theory \cite{Geor}, ${\cal
F}(w)$ coincides with the universal Isgur-Wise function $\xi(w)$
\cite{Isgu,Falk}, which describes the long-distance physics
associated with the light degrees of freedom in the heavy mesons and
is normalized to unity at the zero-recoil point $w=1$. As a
consequence, the normalization of the physical form factor is
determined up to small perturbative corrections and power corrections
of order $(\Lambda_{\rm QCD}/m_Q)^2$ \cite{Luke}, where we use $m_Q$
generically for $m_b$ or $m_c$. Detailed calculations of these
corrections lead to ${\cal F}(1)=0.91\pm 0.03$ \cite{Vcb},
\cite{FaNe}--\cite{Czar}. Therefore, an accurate determination of
$|V_{cb}|$ can be obtained by extrapolating the differential decay
rate to $w=1$.

This analysis has been performed by several experimental groups
\cite{ARGUS}--\cite{DELPHI}. The existing data are compatible with a
linear dependence of the form factor ${\cal F}(w)$ on $w$, with
possible corrections induced by a non-zero curvature. In general, one
may define
\begin{equation}\label{Fexp}
   {\cal F}(w) = {\cal F}(1)\,\Big\{ 1 - \widehat\varrho^2\,(w-1)
   + \widehat c\,(w-1)^2 + O[(w-1)^3] \Big\} \,,
\end{equation}
where $\widehat\varrho$ and $\widehat c$ are referred to as the
charge radius and the convexity, respectively. In order to guide the
extrapolation to zero recoil, it is of interest to gain theoretical
information about these parameters. In the present paper, we
readdress this problem and derive bounds for $\widehat\varrho^2$ and
$\widehat c$ that are much stronger than those obtained in previous
analyses \cite{Taro}--\cite{Boyd2}.

A model-independent method of constraining the $q^2$ dependence of
form factors using analyticity properties of QCD spectral functions
and unitarity was proposed some time ago in
Refs.~\cite{Meim}--\cite{Bour}. More recently, the same method was
applied to the elastic form factor of the $B$ meson
\cite{Taro}--\cite{Boyd1}, which is related by heavy-quark symmetry
to the Isgur-Wise function. The resulting bounds prove to be rather
weak, however, due to the presence of the $\Upsilon$ poles below the
$B\bar B$ threshold. The lack of information about the residues of
these poles, related to the unknown $\Upsilon B\bar B$ couplings,
reduce considerably the constraining power of the method. The
technique of treating, in an optimal way, poles with unknown residues
is to multiply the form factor by specific functions in the complex
$q^2$ plane (the so-called Blaschke factors) with zeros at the
positions of the poles, but with unit modulus along the physical cut.
Since these functions have modulus less than unity below the cut (and
in particular at the zero-recoil point), they weaken the bounds on
the form factor. Hence, the more poles there are below threshold, the
weaker these bounds are. In a recent paper, Boyd et al.\ have applied
this approach to the form factors of interest for $\bar B\to
D^{(*)}\,\ell\,\bar\nu$ decays, i.e.\ to the matrix elements of
flavour-changing heavy-quark currents
\cite{Boyd2}.\footnote{Applications to the heavy-to-light transitions
$\bar B\to\pi,\rho\,\ell\,\bar\nu$ have been considered in
Refs.~\protect\cite{hl1,hl2}.}
However, once again the constraining power of the method is strongly
reduced because of the presence of possible $(\bar b c)$ bound states
below the threshold for $B D^{(*)}$ production. Up to now,
pseudoscalar $B_c$ and vector $B_c^*$ mesons have not been observed
experimentally; they are predicted, however, by potential models
\cite{Eich}--\cite{Chen}. The uncertainty in these model calculations
weakens the derived constraints even further.

Instead of relying on model-dependent predictions about the
properties of $B_c^{(*)}$ mesons, we adopt a different strategy: we
identify a specific $\bar B\to D$ form factor, which does not receive
contributions from the ground-state $B_c$ poles. By applying the
method of Refs.~\cite{Taro}--\cite{Bour}, we derive strong
model-independent constraints on the slope and curvature of this form
factor. Heavy-quark symmetry is then used to relate the form factor
to the function ${\cal F}(w)$ describing $\bar B\to D^*\ell\,\bar\nu$
decays, and to translate the constraints into bounds for the charge
radius $\widehat\varrho$ and the convexity $\widehat c$. These
relations receive symmetry-breaking corrections, which can however be
estimated. They turn out to weaken the bounds only softly, so that
our constraints are stronger than those obtained in previous
analyses. Thus, our results should be used in future determinations
of $|V_{cb}|$. At the end, we briefly consider the corresponding
bounds on the form factor in the decay $\bar B\to D\,\ell\,\bar\nu$.

\section{Derivation of the bounds}

We consider the flavour-changing vector current $V^\mu=\bar
c\,\gamma^\mu b$ and write its matrix element between $\bar B$- and
$D$-meson states in terms of hadronic form factors $F_0(q^2)$ and
$F_1(q^2)$ defined as ($q=p-p'$)
\begin{equation}\label{FFdef}
   \langle D(p')|V^\mu|\bar B(p)\rangle =
   \bigg[ (p+p')^\mu-{m_B^2-m_D^2\over q^2}\,q^\mu \bigg]\,F_1(q^2)
   + {m_B^2-m_D^2\over q^2}\,q^\mu\,F_0(q^2) \,.
\end{equation}
The form factors are real analytic functions in the complex $q^2$
plane, cut along the real axis from the branch point
$t_0=(m_B+m_D)^2$ to infinity. Below the threshold $t_0$, poles can
appear at $q^2=m_{B_c^*}^2$, and also branch points due to
non-resonant states. Their contribution to the form factors can be
obtained using crossing symmetry and the unitarity relation
\begin{equation}\label{unitsum}
   2\,\mbox{Im}\,\langle\,0\,|V^\mu|\bar B(p)\,\bar D(-p')\rangle
   = \sum\!\!\!\!\!\!\!\int\limits_\Gamma\,\,{\rm d}\rho_\Gamma\,
   (2\pi)^4\,\delta^{(4)}(p_\Gamma - q)\,\langle\,0\,|V^\mu
   |\Gamma\rangle\,\langle\Gamma|\bar B(p)\,\bar D(-p')\rangle \,,
\end{equation}
where the sum is over all possible hadron states $\Gamma$ with the
appropriate flavour quantum numbers. The form factor $F_1(q^2)$
receives pole contributions from $B_c^*$ vector
mesons.\footnote{Similarly, the form factors parametrizing the matrix
elements for $\bar B\to D^*$ transitions receive pole contributions
from ground-state pseudoscalar ($B_c$) or vector ($B_c^*$) mesons.}
The scalar form factor $F_0(q^2)$, however, does not couple to
ground-state $B_c$ or $B_c^*$ mesons. We note that $F_0(q^2)$ can, in
principle, receive contributions from scalar $B_c$ states; however,
these are expected to be broad resonances above the branch point due
to two-particle intermediate states of the form $(B_c^{(*)}+h)$,
where $h$ is a light hadron. We shall discuss these sub-threshold
singularities below. Assuming for the moment that their effect is
negligible, the form factor $F_0(q^2)$ can be considered a real
analytic function in the complex $q^2$ plane except for the cut from
$t_0$ to infinity.

Consider then the vacuum polarization tensor
\begin{eqnarray}
   \Pi^{\mu\nu}(q) &=& i\int\!\mbox{d}^4x\,e^{iq\cdot x}\,
    \langle\,0\,|\,T\{V^\mu(x),V^{\dagger\nu}(0)\}|\,0\,\rangle
    \nonumber\\
   &=& (q^\mu q^\nu - g^{\mu \nu} q^2)\,\Pi(q^2)
    + g^{\mu \nu} D(q^2) \,.
\end{eqnarray}
The invariant amplitudes $\Pi(q^2)$ and $D(q^2)$ satisfy
once-subtracted dispersion relations, so it is convenient to consider
their first derivatives with respect to $q^2$. Applying a unitarity
relation similar to (\ref{unitsum}) to the corresponding spectral
functions, we obtain the positivity conditions ($t=q^2$)
\begin{equation}
   \mbox{Im}\,\Pi(t+i\epsilon )\ge 0 \,,\qquad
   \mbox{Im}\,D(t+i\epsilon )\ge 0 \,;\qquad t\ge t_0 \,.
\end{equation}
Therefore, if we retain in the unitarity sum only the contribution of
the two-particle state $|\bar B\,\bar D\rangle$, we obtain rigorous
lower bounds on the spectral functions. Being interested in the form
factor $F_0(t)$, we retain the inequality for the longitudinal
amplitude $D(t)$, which reads
\begin{equation}
   \mbox{Im}\,D(t+i\epsilon)\ge {n_f\over 16\pi t^2}\,t_0\,t_1
   \sqrt{(t-t_0)(t-t_1)}\,|F_0(t)|^2 \,,
\end{equation}
where
\begin{equation}
   t_0 = (m_B + m_D)^2 \,, \qquad t_1 = (m_B - m_D)^2 \,,
\end{equation}
and $n_f$ is the number of light flavours. The factor of $n_f$
appears since SU$(n_f)$ light-flavour multiplets of heavy-meson
states contribute with the same weight to the unitarity sum. Below we
will take $n_f=2.5$ to account for the breaking of flavour symmetry
by the mass of the strange quark. Using the above inequality in the
dispersion relation
\begin{equation}
   D'(q^2) = {1\over\pi} \int\limits_{t_0}^\infty\!\mbox{d}t\,
   {\mbox{Im}\,D(t+i\epsilon)\over(t-q^2)^2} \,,
\end{equation}
we obtain the inequality
\begin{equation}\label{ineq}
   {n_f\over 16\pi^2}{t_0\,t_1\over D'(q^2)}
   \int\limits_{t_0}^\infty\!\mbox{d}t\,
   {\sqrt{(t-t_0)(t-t_1)}\over t^2(t-q^2)^2}\,|F_0(t)|^2 \le 1 \,.
\end{equation}
On the other hand, if $q^2\ll(m_b+m_c)^2$, the quantity $D'(q^2)$ can
be calculated using the operator product expansion. The leading-order
expression is
\begin{equation}\label{Dfun}
   D'(q^2) = {N_c\over 4\pi^2} \int\limits_0^1\!\mbox{d}x\,x(1-x)\,
   {x m_b^2 + (1-x) m_c^2 - m_b m_c\over
    x m_b^2 + (1-x) m_c^2 - x(1-x) q^2} \,,
\end{equation}
where $m_b$ and $m_c$ are the heavy-quark masses. The choice of the
value of $q^2$ will be discussed below. We note that, for $q^2=0$,
one obtains the simple result
\begin{equation}
   D'(0) = {N_c\over 24\pi^2}\,\Bigg\{
   {(1-4r+r^2)(1+r+r^2)\over(1-r^2)^2} - {12 r^3\ln r\over(1-r^2)^3}
    \Bigg\} \,,
\end{equation}
where $r=m_c/m_b$. Expressions for the order-$\alpha_s$ corrections
to the polarization functions in the case of unequal quark masses
have been derived in Refs.~\cite{Broa,Gene}. The leading
non-perturbative contribution is proportional to the gluon condensate
and, for dimensional reasons, suppressed by four powers of a large
mass scale. As in the case of QCD sum rules for the charmonium and
bottomonium systems \cite{SVZ}, this contribution is very small.
Below, we shall include possible effects of all these corrections in
a very conservative way.

It is convenient to perform the conformal mapping
\begin{equation}\label{mapp}
   z = {\sqrt{t_0-t}-\sqrt{t_0-t_1}\over\sqrt{t_0-t}+\sqrt{t_0-t_1}}
   \,,
\end{equation}
which transforms the cut $t$-plane onto the interior of the unit disc
$|z|<1$, such that the point $t_1$ is mapped onto the origin $z=0$,
while the branch point $t_0$ is mapped onto $z=-1$. Applying
standard techniques in the theory of analytic functions \cite{math1},
we write the inequality (\ref{ineq}) as
\begin{equation}\label{phiF0}
   {1\over 2\pi} \int\limits_0^{2\pi}\!\mbox{d}\theta\,
   |\phi(e^{i\theta})|^2\,|\tilde F(e^{i\theta})|^2 \leq 1 \,,
\end{equation}
where $\tilde F(z)\equiv F_0(t(z))$. The function $\phi(z)$ is
analytic and without zeros inside the unit disc, and its modulus
squared on the boundary of the unit disc is equal to the weight
function appearing in (\ref{ineq}) multiplied by the Jacobian of the
conformal transformation. In general
\begin{equation}
   \ln\phi(z) = {1\over 4\pi} \int\limits_0^{2\pi}\!\mbox{d}\theta\,
   {e^{i\theta}+z\over e^{i\theta}-z}\,\ln|\phi(e^{i\theta})|^2 \,.
\end{equation}
In the present case, a straightforward calculation gives \cite{Bour}:
\begin{equation}\label{phifun}
   \phi(z) = \phi(0)\,
   {(1+z) (1-z)^{3/2}\over(1-d_1 z)^2 (1-d_2 z)^2} \,,
\end{equation}
where
\begin{eqnarray}
   \phi(0) &=& \sqrt{n_f\over 2\pi D'(q^2)}\,
    {\sqrt{t_0\,t_1}\over t_0-t_1}\,{(1-d_1)^2 (1-d_2)^2\over 16}
    \,, \nonumber\\
   d_1 &=& {\sqrt{t_0}-\sqrt{t_0-t_1}\over\sqrt{t_0}+\sqrt{t_0-t_1}}
    \,, \nonumber\\
   d_2 &=& {\sqrt{t_0-q^2}-\sqrt{t_0-t_1}
          \over\sqrt{t_0-q^2}+\sqrt{t_0-t_1}} \,.
\end{eqnarray}

The inequality (\ref{phiF0}) is a condition for the norm on the
Hilbert space $H^2$ of analytic functions, which implies rigorous
constraints for the values of the function $\tilde F(z)$ and its
derivatives at interior points \cite{math1}. We shall transform these
constraints into bounds on the slope and curvature of the function
\begin{equation}
   f_0(w) = f_0(1)\,\Big\{ 1 - \varrho_0^2\,(w-1)
   + c_0\,(w-1)^2 + O[(w-1)^3] \Big\} \,,
\end{equation}
which is related to the original form factor $F_0(q^2)$ by
\begin{equation}\label{f0def}
   F_0(q^2) = {m_B+m_D\over 2\sqrt{m_B m_D}}\,\bigg[
   1 - {q^2\over(m_B+m_D)^2} \bigg]\,f_0(w(q^2)) \,.
\end{equation}
The relation between $w$ and $q^2$ is given by (\ref{wdef}) with
$m_D^*$ replaced by $m_D$. The definition (\ref{f0def}) is such that
in the heavy-quark limit the function $f_0(w)$ coincides with the
Isgur-Wise form factor \cite{NeRi}. The zero-recoil point $w=1$
corresponds to $q^2=t_1$, i.e.\ to $z=0$. The parameters $f_0(1)$,
$\varrho_0^2$ and $c_0$ are related to the function $\tilde F(z)$ and
its derivatives with respect to $z$ through
\begin{eqnarray}
   \tilde F(0) &=& \beta\,f_0(1) \,, \nonumber\\
   {\tilde F'(0)\over\tilde F(0)} &=& 4 - 8\varrho_0^2 \,,
    \nonumber\\
   {\tilde F''(0)\over\tilde F(0)} &=& 128 c_0 - 96\varrho_0^2
    + 16 \,,
\end{eqnarray}
where
\begin{equation}
   \beta = {2\sqrt{m_B m_D}\over m_B+m_D} \simeq 0.879 \,.
\end{equation}
Using well-known results in the interpolation theory for the Hilbert
space $H^2$ \cite{math1}, we obtain from (\ref{phiF0}) the inequality
\begin{equation}\label{phiFeq}
   \big[(\phi\,\tilde F)(0)\big]^2 + \big[(\phi\,\tilde F)'(0)\big]^2
   + {1\over 4}\,\big[(\phi\,\tilde F)''(0)\big]^2 < 1 \,.
\end{equation}
Written in terms of $\varrho_0^2$ and $c_0$, this becomes the
equation for an ellipse:
\begin{equation}\label{ellipse}
   (\varrho_0^2-\rho^2)^2 + 64\,\Big[ (c_0-k)
   - \chi\,(\varrho_0^2-\rho^2) \Big]^2 < K^2 \,,
\end{equation}
where
\begin{eqnarray}\label{pars}
   \rho^2 &=& {7\over 16} + {d_1+d_2\over 4} \,, \nonumber\\
   \chi &=& {11\over 16} + {d_1+d_2\over 4} \,, \nonumber\\
   k &=& {115\over 512} + {11\over 64}\,(d_1+d_2)
    + {d_1^2 + 4 d_1 d_2 + d_2^2\over 64} \,, \nonumber\\
   K &=& {1\over 8}\sqrt{{1\over\big[\phi(0)\,\beta\,f_0(1)\big]^2}
    - 1} \,.
\end{eqnarray}
{}From (\ref{ellipse}), one can derive strict upper and lower bounds
on the charge radius and the convexity of the form factor $f_0(w)$,
which read
\begin{eqnarray}\label{interv}
   \rho^2 - K &<& \varrho_0^2 < \rho^2 + K \,, \nonumber\\
   k - \sqrt{\textstyle{1\over 64} + \chi^2}\,K &<& c_0 <
    k + \sqrt{\textstyle{1\over 64} + \chi^2}\,K \,.
\end{eqnarray}
Note that the large numerical factor 64 in front of the second term
in (\ref{ellipse}) implies a strong correlation between the slope
parameter $\varrho_0^2$ and the curvature $c_0$ (since $K^2\ll 1$,
see below). In other words, the resulting ellipse in the
$\varrho_0^2$--$c_0$ plane is almost degenerate to a line, and to a
good approximation
\begin{equation}\label{approx}
   c_0 \simeq \chi\varrho_0^2 + (k-\chi\rho^2) \,.
\end{equation}
It is remarkable that the parameters in this relation only depend on
the meson masses and $q^2$; the dynamical information encoded in
$\phi(0)$ and $f_0(1)$ does not enter here (once these parameters are
such that $K^2\ll 64$).

Let us comment, at this point, on an obvious extension of our
approach. Instead of (\ref{phiFeq}), one could write a more general
inequality involving higher derivatives of the product $\phi\,\tilde
F$. This would further constrain the form factor near zero recoil. We
refrain from presenting such an extension because of two
approximations inherent in our treatment: first, sub-threshold
singularities give an increasingly larger contribution to higher
derivatives of $\phi\,\tilde F$ (see eq.~(\ref{deltas}) below);
secondly, the relations among the form factors implied by heavy-quark
symmetry, which will be used later to derive from our results bounds
on the form factors of interest in semileptonic $B$ decays, are
expected to break down for higher derivatives of the form factors.

\begin{figure}[htb]
   \epsfxsize=9cm
   \centerline{\epsffile{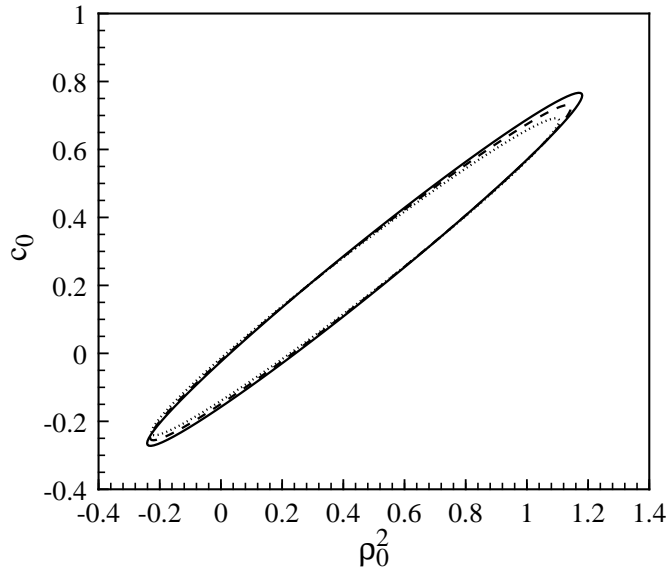}}
   \centerline{\parbox{14cm}{\caption{\label{fig:ellipse}
Allowed regions in the $\varrho_0^2$--$c_0$ plane for different
values of $q^2$.
   }}}
\end{figure}

For the numerical evaluation of the bounds, we first consider the
case where $q^2=0$, so that
\begin{equation}
   d_1 = d_2 = \bigg(
   {\sqrt{m_B}-\sqrt{m_D}\over\sqrt{m_B}+\sqrt{m_D}} \bigg)^2
   \simeq 0.065 \,.
\end{equation}
This leads to $\rho^2\simeq 0.470$, $\chi\simeq 0.720$ and $k\simeq
0.247$. The only model-dependent quantity is the ``radius'' $K$,
which depends on the product $\phi(0)\,f_0(1)$. Using
$r=m_c/m_b=0.29\pm 0.03$ for the ratio of the heavy-quark masses, as
well as $n_f=2.5$, we find $\phi(0)=0.282\pm 0.018$. As discussed
above, QCD corrections are expected to modify this result in a
moderate way. Moreover, in the heavy-quark limit we have (including
short-distance corrections) $f_0(1)\simeq 1.02$ \cite{review}, and
corrections to this result are of order $(\Lambda_{\rm QCD}/m_Q)^2$
and should not exceed 10\%. Thus, we believe it is conservative to
assume that the total uncertainty in the product $\phi(0)\,f_0(1)$ is
less than 30\%. As our goal is to derive bounds, we take the smallest
possible value, i.e.\ $\phi(0)\,f_0(1)=0.20$, which leads to the
largest value of $K$ and thus to the largest allowed domain in the
$\varrho_0^2$--$c_0$ plane. With this choice, we obtain $K\simeq
0.70$. According to (\ref{interv}), the allowed intervals for the
slope and curvature are then given by
\begin{equation}
   -0.23 < \varrho_0^2 < 1.17 \,, \qquad -0.26 < c_0 < 0.76 \,,
\end{equation}
and the approximate relation (\ref{approx}) between the two
parameters reads
\begin{equation}
   c_0 \simeq 0.72\varrho_0^2 - 0.09 \,.
\end{equation}
The corresponding narrow ellipse is shown by the solid curve in
Fig.~\ref{fig:ellipse}. The allowed region can be reduced further by
considering values $q^2>0$; however, the QCD calculation is reliable
only if $q^2\ll(m_b+m_c)^2$. As an example, we show by the dashed and
dotted curves the two ellipses obtained for $q^2=10$ and 20~GeV$^2$.
The gain that can be achieved in this way is rather small, and to be
conservative we shall take $q^2=0$ from now on.

\section{Sub-threshold singularities}

A last question that has to be discussed is the effect of possible
sub-threshold singularities. These are due to states of the form
$(B_c^{(*)}+h)$, where $h$ is a light hadron, or to scalar $B_c$
resonances, which contribute to the unitarity sum in (\ref{unitsum}).
Because these scalar mesons are predicted to have masses that lie
above the branch point due to two-particle intermediate states
\cite{Eich}--\cite{Chen}, we shall treat them as resonances in these
two-particle channels. The lowest intermediate states in the
unitarity relation for the form factor $F_0(t)$ are then given by
$(B_c^{(*)}+\pi)$. However, these contributions are suppressed by
isospin symmetry, so the lowest states which contribute significantly
are $(B_c^{(*)}+\eta, \omega)$. Consequently, the sub-threshold cut
is rather short. We parametrize its contribution to the discontinuity
of the form factor by (a similar parametrization was adopted in
Ref.~\cite{Boyd2}):
\begin{equation}
   \mbox{Im}\,F_0(t+i\epsilon) = {C\over\sqrt{t_0}}\,\sqrt{t-t_+}
   \,,\quad t_+\le t\le t_0 \,,
\end{equation}
where $t_+=(m_{B_c}+m_\eta)^2$, and $C$ is a dimensionless quantity
expected to be of order unity. We estimated this quantity in a model
where the form factors and amplitudes in the unitarity integral in
(\ref{unitsum}) are saturated by the nearest resonances, and find
that $C$ is a slowly varying quantity in the interval $t_+<t<t_0$,
which takes values in the range 0.5--3.\footnote{We note that the
contribution of the sub-threshold cut is suppressed by the OZI rule,
since the couplings $B_c B_c^* h$ appearing in $\mbox{Im}\,F_0$ are
described by ``hair-pin diagrams''.}

Once an approximation of $\mbox{Im}\,F_0$ is adopted, the
effect of sub-threshold singularities can be treated in an exact way
\cite{Irin}. In this case, the function $\tilde F(z)$ obtained after
the conformal mapping (\ref{mapp}) is no longer analytic. Yet, the
formalism presented above can be applied to the new function
\begin{equation}\label{gdef}
   g(z) = \phi(z)\tilde F(z) - {1\over\pi}
   \int\limits_{-1}^{z_+}\!\mbox{d}x\,
   {\phi(x)\,\mbox{Im}\,\tilde F(x)\over x-z} \,,
\end{equation}
where $z_+\equiv z(t_+)$ is the position of the lowest sub-threshold
branch point. We have used the fact that the function $\phi(z)$ is
real analytic inside the unit disc. By definition, the function
$g(z)$ is analytic for $|z|<1$, since the discontinuity of the
product $\phi(z)\tilde F(z)$ is compensated by the subtraction term.
Substituting (\ref{gdef}) into the inequality (\ref{phiF0}), we
obtain
\begin{equation}
   {1\over 2\pi} \int\limits_0^{2\pi}\!\mbox{d}\theta\,\Bigg|\,
   g(e^{i\theta}) + {1\over \pi}\int\limits_{-1}^{z_+}\!\mbox{d}x\,
   {\phi(x)\,\mbox{Im}\,\tilde F(x)\over x-e^{i\theta}} \Bigg|^2
   \le 1 \,.
\end{equation}
By performing a Fourier expansion of the integrand, which contains
now both positive- and negative-frequency Fourier coefficients, and
taking into account the orthogonality of these coefficients, we
obtain, after a straightforward calculation, the following inequality
instead of (\ref{phiFeq}):
\begin{eqnarray}\label{deltas}
   &&\Big[ \phi(0)\tilde F(0) - \delta_0 \Big]^2
    + \Big[ (\phi\,\tilde F)'(0) - \delta_1 \Big]^2
    + {1\over 4}\,\Big[ (\phi\,\tilde F)''(0) - \delta_2 \Big]^2
    \nonumber\\
   &&\mbox{}< 1 - {1\over\pi^2}\int\limits_{-1}^{z_+}\!\mbox{d}x
    \int\limits_{-1}^{z_+}\!\mbox{d}y\,
    {\phi(x)\phi(y)\,\mbox{Im}\,\tilde F(x)\,\mbox{Im}\,\tilde F(y)
    \over 1 - xy} \equiv 1 - \delta_{xy} \,,
\end{eqnarray}
where
\begin{equation}
   \delta_n = {n!\over\pi} \int\limits_{-1}^{z_+}\!
   \mbox{d}x\,{\phi(x)\,\mbox{Im}\,\tilde F(x)\over x^{n+1}} \,.
\end{equation}
This inequality shows the corrections induced by the additional
branch point below the threshold $t_0$.

The net effect is that the parameters describing the shape of the
ellipse in (\ref{ellipse}) are modified. Instead of $\rho^2$, $k$ and
$K$ in (\ref{pars}), we now obtain new parameters
$\rho^2+\delta\rho^2$, $k+\delta k$ and $\sqrt{K^2+\delta K^2}$,
where
\begin{eqnarray}
   \delta\rho^2 &=& - {\delta_1\over 8N} \,, \nonumber\\
   \delta k &=& - \bigg( {11\over 128} + {d_1 + d_2\over 32} \bigg)\,
    {\delta_1\over N} + {\delta_2\over 128 N} \,, \nonumber\\
   \delta K^2 &=& {1\over 64}\,\bigg[ {2\delta_0\over N}
    - {\delta_0^2 + \delta_{xy}\over N^2} \bigg] \,,
\end{eqnarray}
and $N=\phi(0)\,\beta\,f_0(1)$. Numerically, we find that
\begin{eqnarray}
   \delta\rho^2 &\simeq& -4.0\times 10^{-3}\,C \,, \nonumber\\
   \delta k &\simeq& -2.1\times 10^{-3}\,C \,, \nonumber\\
   \delta K^2 &\simeq& -0.7\times 10^{-3}\,C - 10^{-5} C^2 \,.
\end{eqnarray}
For $C$ of order unity, these corrections have a negligible effect on
the bounds. Note, in particular, that the parameter $\chi$ remains
unchanged, and that the combination $(k-\chi\rho^2)$ changes by only
$0.8\times 10^{-3}\,C$. Thus, the approximate linear relation
(\ref{approx}) is essentially unaffected by sub-threshold
singularities.

\section{Bounds on $\bar B\to D^{(*)}\ell\,\bar\nu$ form factors}

Our final task is to use heavy-quark symmetry to translate the bounds
and constraints derived above into bounds for the slope and curvature
of the form factor ${\cal F}(w)$, which describes the recoil spectrum
in the decay $\bar B\to D^*\ell\,\bar\nu$. Our definition of the
parameters $\varrho_0^2$ and $c_0$ was such that they agree, in the
heavy-quark limit, with the slope $\widehat\varrho^2$ and the
curvature $\widehat c$ of the physical form factor ${\cal F}(w)$.
Differences are induced, however, by corrections that break the
heavy-quark symmetry. In general, there are perturbative corrections
of order $\alpha_s(m_Q)$, and power corrections of order
$\Lambda_{\rm QCD}/m_Q$. A calculation of the latter requires
non-perturbative techniques such as lattice field theory; this is
beyond the scope of this paper. Here we shall include the
short-distance corrections, which can be calculated in a
model-independent way using perturbation theory.

\begin{figure}[htb]
   \epsfxsize=9cm
   \centerline{\epsffile{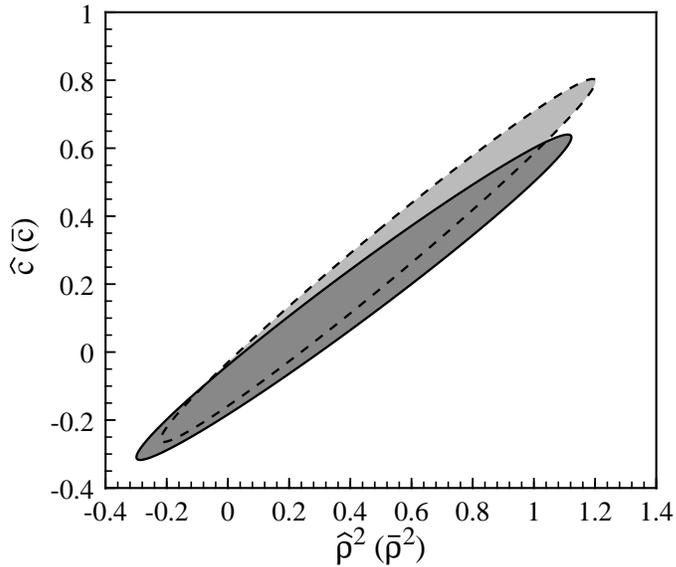}}
   \centerline{\parbox{14cm}{\caption{\label{fig:physical}
Allowed regions for the slope and curvature of the form factors
${\cal F}(w)$ (dark-shaded area) and ${\cal G}(w)$ (light-shaded
area), which describe $\bar B\to D^{(*)}\ell\,\bar\nu$ decays.
   }}}
\end{figure}

To derive the relations between the slope and curvature parameters of
different form factors we use the results of
Refs.~\cite{QCD,Pasc,review}, where the relations between heavy-meson
form factors have been calculated including one-loop QCD corrections.
Expanding the corresponding lengthy expressions around the
zero-recoil point $w=1$, we find ($r=m_c/m_b$)
\begin{equation}
   \widehat\varrho^2 = \varrho_0^2 - {4\alpha_s\over 9\pi}\,
   \Big[ 1 - 4\psi(r) \Big] + O(\Lambda_{\rm QCD}/m_Q)
\end{equation}
for the relation between the slope parameters, and
\begin{eqnarray}
   \widehat c &=& c_0 - {2\alpha_s\over 27\pi}\,\Bigg[
    {7-2 r+7 r^2\over(1-r)^2} + {1+42 r+r^2\over(1-r)^2}\,\psi(r)
    \Bigg] \nonumber\\
   &&\mbox{}- {4\alpha_s\over 9\pi}\,\varrho_0^2\,
    \Big[ 1 - 4\psi(r) \Big]    + O(\Lambda_{\rm QCD}/m_Q)
\end{eqnarray}
for the relation between the curvatures, where
\begin{equation}
   \psi(r) = {r\over(1-r)^2}\,\bigg(
   {1+r\over 1-r}\,\ln r + 2 \bigg) \,.
\end{equation}
To evaluate these expressions, we use $r=0.29$ and
$\alpha_s=\alpha_s(\sqrt{m_b m_c})=0.26$, leading to
\begin{equation}
   \widehat\varrho^2 \simeq \varrho_0^2 - 0.06 \,, \qquad
   \widehat c \simeq c_0 - 0.06 - 0.06\varrho_0^2 \,.
\end{equation}
The corresponding ellipse is shown by the dark-shaded area in
Fig.~\ref{fig:physical}. Note that the boundary of the allowed region
is uncertain by an amount of order $\Lambda_{\rm QCD}/m_Q$. Given
this fact, we can safely replace the ellipse with the approximate
linear relation between the curvature and the slope parameter
analogous to (\ref{approx}). This relation is the central result of
our analysis:
\begin{equation}
   \widehat c\simeq 0.66\widehat\varrho^2 - 0.11
   + O(\Lambda_{\rm QCD}/m_Q) \,.
\end{equation}
The average experimental value of the slope parameter, as extracted
from a linear fit to the recoil spectrum, is
$\widehat\varrho^2=0.82\pm 0.09$
\cite{ARGUS}--\cite{DELPHI},\cite{Dubna}. Allowing for a positive
curvature of the form factor, the true slope parameter may be
somewhat larger; however, the experimental value is already close to
the upper bound for $\widehat\varrho^2$ in Fig.~\ref{fig:physical}.
Using this value, we predict that
\begin{equation}
   \widehat c \gsim 0.43\pm 0.06 + O(\Lambda_{\rm QCD}/m_Q) \,.
\end{equation}
We thus expect a moderate positive curvature of the form factor
${\cal F}(w)$.

Let us, for completeness, also consider the semileptonic decay $\bar
B\to D\,\ell\,\bar\nu$. The differential decay rate for this process
is given by \cite{Vcb}
\begin{equation}
   {\mbox{d}\Gamma(\bar B\to D\,\ell\,\bar\nu)\over\mbox{d}w}
   = {G_F^2\over 48\pi^3}\,(m_B+m_D)^2\,m_D^3\,(w^2-1)^{3/2}\,
   |V_{cb}|^2\,{\cal G}^2(w) \,,
\end{equation}
where $w=v_B\cdot v_D$, and the hadronic form factor ${\cal G}(w)$
obeys an expansion similar to (\ref{Fexp}):
\begin{equation}
   {\cal G}(w) = {\cal G}(1)\,\Big\{ 1 - \bar\varrho^2\,(w-1)
   + \bar c\,(w-1)^2 + O[(w-1)^3] \Big\} \,.
\end{equation}
In this case, the normalization at zero recoil is known only up to
first-order power corrections \cite{NeRi}. An explicit calculation
using the QCD sum-rule approach leads to ${\cal G}(1)/{\cal
F}(1)=1.08\pm 0.06$ \cite{LNN}, which can be combined with ${\cal
F}(1)=0.91\pm 0.03$ to give ${\cal G}(1)=0.98\pm 0.07$. Thus, in
principle, a measurement of the recoil spectrum in $\bar B\to
D\,\ell\,\bar\nu$ decay provides an independent determination of
$|V_{cb}|$ with a theoretical accuracy not much less than in the
decay $\bar B\to D^*\ell\,\bar\nu$. For the slope and curvature of
the form factor ${\cal G}(w)$, we obtain the relations
\begin{eqnarray}
   \bar\varrho^2 &=& \varrho_0^2 + {4\alpha_s\over 9\pi}\,
    \Big[ 1 + 3\psi(r) \Big]
    + O(\Lambda_{\rm QCD}/m_Q) \,, \nonumber\\
   \bar c &=& c_0 + {4\alpha_s\over 45\pi}\,\Bigg[
    {2-9 r+2 r^2\over(1-r)^2} - {30 r\over(1-r)^2}\,\psi(r)
    \Bigg] \nonumber\\
   &&\mbox{}+ {4\alpha_s\over 9\pi}\,\varrho_0^2\,
    \Big[ 1 + 3\psi(r) \Big] + O(\Lambda_{\rm QCD}/m_Q)  \,.
\end{eqnarray}
Numerically, this gives $\bar\varrho^2\simeq\varrho_0^2+0.02$ and
$\bar c\simeq c_0+0.01+0.02\varrho_0^2$. The corresponding ellipse
is shown by the light-shaded area in Fig.~\ref{fig:physical}. The
approximate linear relation is
\begin{equation}
   \bar c\simeq 0.74\bar\varrho^2 - 0.09
   + O(\Lambda_{\rm QCD}/m_Q) \,.
\end{equation}

\section{Conclusion}

Using analyticity, unitarity and heavy-quark symmetry, we have
derived from QCD conservative bounds on the slope and curvature of
the form factors describing the semileptonic decays $\bar B\to
D^{(*)}\ell\,\bar\nu$. The allowed regions for these parameters are
displayed in Fig.~\ref{fig:physical}. Our method employs heavy-quark
symmetry in such a way as to avoid the problem of sub-threshold poles
due to ground-state $B_c$ mesons. Thus, our bounds are stronger than
the ones obtained in previous analyses. In particular, we find that
the curvature $\widehat c$ and the slope $\widehat\varrho^2$ of the
form factor ${\cal F}(w)$ governing the decay $\bar B\to
D^*\ell\,\bar\nu$ are related by $\widehat c\simeq
0.66\,\widehat\varrho^2-0.11$, and that $\widehat\varrho^2<1.11$.
We propose to use the first of these relations in future
determinations of $|V_{cb}|$ from the recoil spectrum in this decay.
It allows the linear fit of this spectrum to be extended to a
quadratic one without introducing a new parameter.

\vspace{0.3cm}
{\it Acknowledgements:\/}
We would like to thank C.T.~Sachrajda for useful discussions. One of
us (I.C.) is grateful to the CERN Theory Division and the ATLAS
Collaboration for support and hospitality during her stay at CERN,
and to M.C.~Gonzalez-Garcia and F.~Cotorobai for help in using the
CERN computing system.


\end{document}